\documentclass[aps,pre,reprint]{revtex4-2}

\usepackage[utf8]{inputenc}

\usepackage{amsmath}
\usepackage{amssymb}
\usepackage{amsfonts}
\usepackage{graphicx} 
\usepackage{hyperref} 
\usepackage{xcolor}
\usepackage[normalem]{ulem}

\begin{document}

\title{A Critical Assessment of the Brain Criticality Hypothesis}

\author{Chesson Sipling}
\email{email: csipling@ucsd.edu}
\affiliation{Department of Physics, University of California San Diego, La Jolla, CA 92093}

\author{Yuan-Hang Zhang}
\email{email: yuz092@ucsd.edu}
\affiliation{Department of Physics, University of California San Diego, La Jolla, CA 92093}

\author{Massimiliano Di Ventra}
\email{email: diventra@physics.ucsd.edu}
\affiliation{Department of Physics, University of California San Diego, La Jolla, CA 92093}

\begin{abstract}

A major unresolved question in Neuroscience is: What is the origin of the observed scale-invariant correlations in neural activity? Many researchers support the ``criticality hypothesis,'' which proposes that the brain operates near criticality, optimizing various information processing functions. However, the nature and behavior of criticality in cortical systems are still unclear. Alternatively, this opinion paper highlights that the coupling between neurons and slowly varying energetic resources, which may act as a form of ``memory,'' alone may be sufficient to generate a robust phase of neural activity with scale-invariant correlations. This {\it memory-induced long-range order} phase could provide a more natural explanation of the existing experimental data than the criticality hypothesis.

\end{abstract}

\maketitle




\section{What is the Criticality Hypothesis?}

\begin{figure}
\centering
\includegraphics[width=0.75\linewidth]{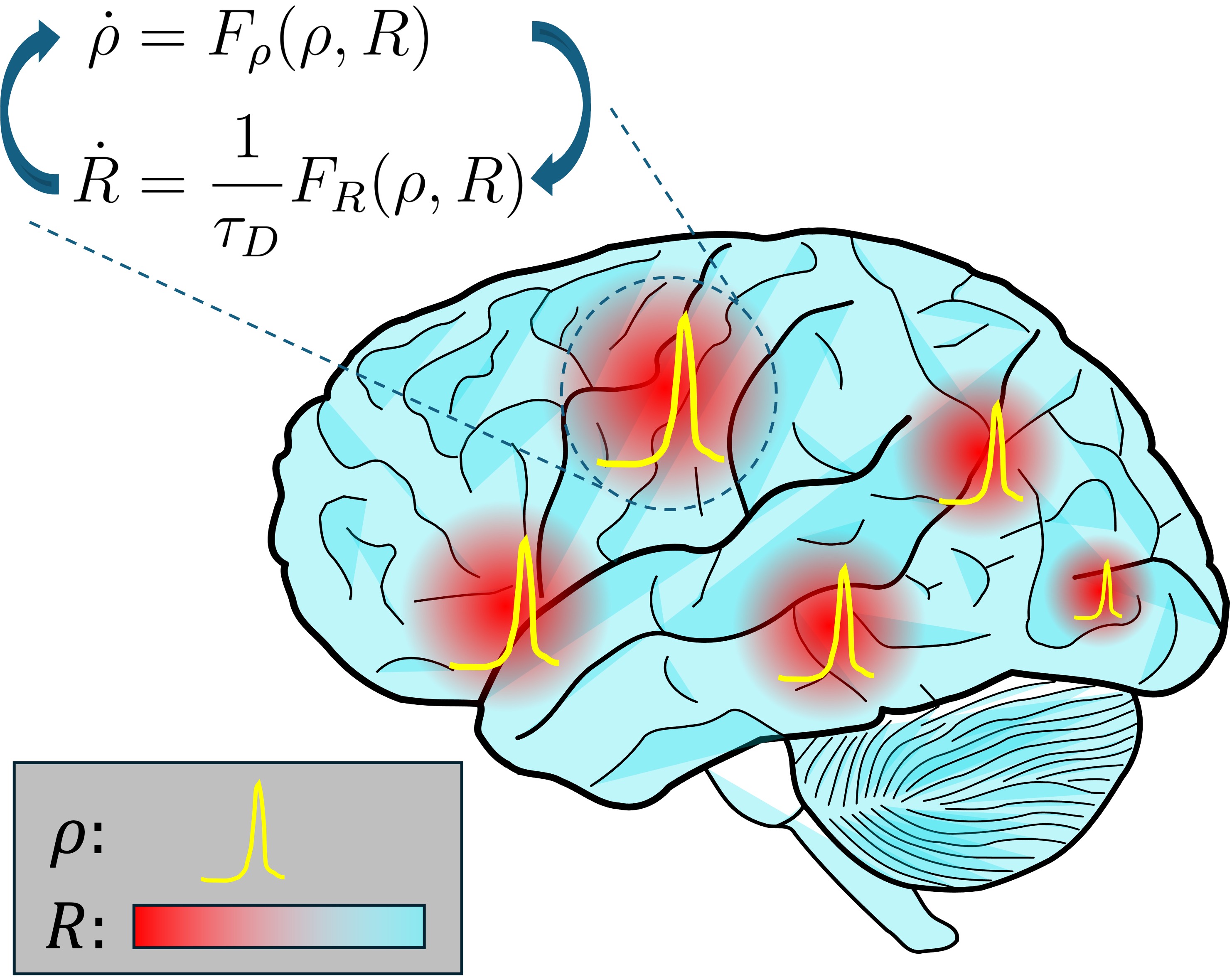}
\caption{A sketch of the human brain. Here, two fundamental dynamical variables in brain behavior are considered in a conceptual schematic: the degree of electrical activity, $\rho$, and the amount of available resources that enable such activity, $R$. Bright yellow spikes represent bursts of electrical activity in neurons. The background colors schematically represent the degree of available resources, with a lower value indicated by red and a higher value indicated by blue. The explicit coupling between these variables is highlighted by the generic dynamical equations at the top of the figure; cf. Eqs.~(\ref{rho}) and~(\ref{R}). Note that the resources are most depleted in regions where neural activity has recently peaked.}
\label{brain_fig}
\end{figure}

The basic mechanisms of neural communication are well-established: {\bf neurons} (see glossary) communicate through electrical impulses called {\bf spikes}, which consume metabolic resources like Adenosine Triphosphate (ATP). At {\bf synapses}---the junctions between neurons---these electrical signals trigger the release of chemical {\bf neurotransmitters} that either excite or inhibit spiking in neighboring cells. This relatively simple, local interaction can generate large-scale cascades of activity throughout the brain~\cite{orlandi2013noise, yaghoubi2018neuronal}. Even when framing at the mesoscopic scale, considering the level of activity and synaptic resources in larger regions of neural tissue (see Fig.~\ref{brain_fig}), large-scale structures emerge~\cite{Beggs11167, 10.1371/journal.pcbi.1000846, di2018landau, sun2025}. Despite this mechanistic understanding, a fundamental question remains: how do local interactions give rise to the complex, collective dynamics observed in neural systems?

\begin{figure}
\centering
\includegraphics[width=0.75\linewidth]{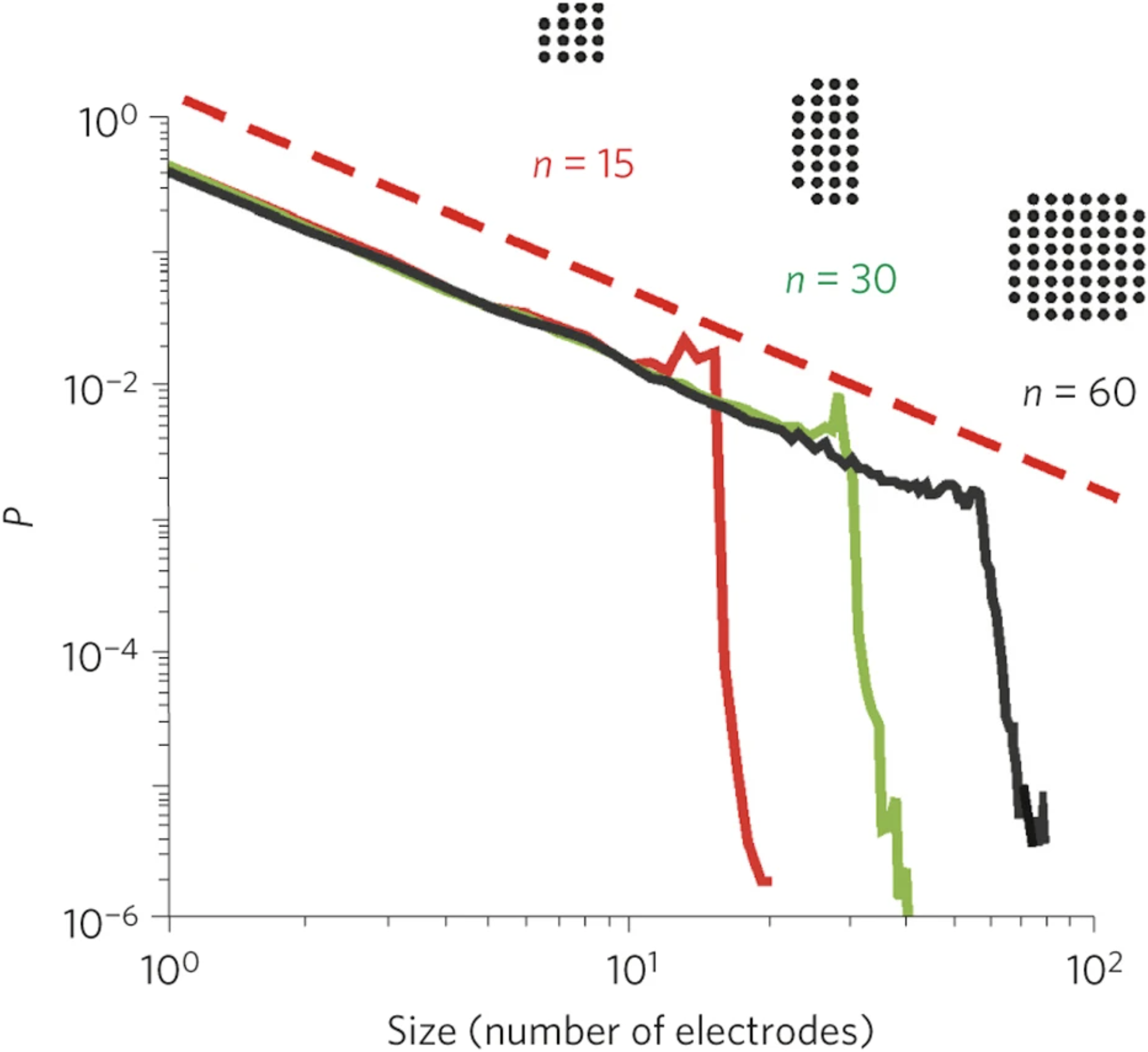}
\caption{Avalanche size, $s$, distributions extracted from electrode arrays on cultured rat cortices. Arrays of size 15, 30, and 60 were used, and a clear power-law decay in the number of activated electrodes is observed up to the size of each array. This is consistent with scale-invariant behavior over the observed range, with a cutoff set by array size. The dotted red line indicates a power-law decay of $s^{-3/2}$. Original visualization from Beggs and Plenz~\cite{Beggs11167}, updated version adapted from Chialvo~\cite{chialvo2010emergent}.}
\label{beggs_experiment_fig}
\end{figure}

One striking example of these collective dynamics appears in {\bf neuronal avalanches}, the observed bursts in electrical activity when several neurons or regions of neural activity fire in quick succession. Studies have found that the probability of an avalanche of a certain size, $P(s)$, occurring (i.e., with a certain number of spikes involved) decays approximately as a power law with avalanche size, $P(s)\sim s^{\alpha}$, with $\alpha$ some number (see Fig.~\ref{beggs_experiment_fig})~\cite{Beggs11167, 10.1371/journal.pcbi.1000846}. This means that there is no single characteristic avalanche size in the brain; while small bursts of activity occur more frequently, larger ones are still expected, given sufficient time. In other words, large bursts of activity are not rare events.

As one explanation for these experimental data, researchers have proposed the so-called {\bf criticality hypothesis}~\cite{10.1098/rsta.2007.2092, PhysRevLett.108.208102, 10.1162/netn_a_00269}. It asserts that the brain naturally operates near {\bf criticality: specifically, near the interface of a continuous transition between phases of activity}. Here, a ``phase'' of activity refers to a distinct mode or type of characteristic behavior (e.g., oscillations, synchronization, etc.). For example, there is a critical point at the ferromagnetic-paramagnetic {\bf phase transition}. At these interfaces, several observables decay as power laws, and near them, this power-law decay is approximate. Hence, this could explain the detected avalanche distributions. However, this hypothesis may not fully explain all currently available experimental data. This, along with several other experimental and analytical considerations, has made the criticality hypothesis a source of heated debate~\cite{chialvo2010emergent, 10.3389/fphys.2012.00163, 10.3389/fncom.2022.703865, 10.1371/journal.pcbi.1000846, DestexheENEURO.0551-20.2021, criticality_setpoint_review, how_critical_is_brain_criticality, 10.3389/fnsys.2014.00166}.

Importantly, as has already been anticipated, neurons require resources (e.g., ATP, at the microscopic scale) to fire. These resources act on much slower time scales compared to the firing of the neurons, hence providing some sort of {\bf memory} to the neurons. Recent research~\cite{sun2025} suggests that this memory promotes a {\it phase} of power-law correlations in the neural activity, thus providing a natural explanation for the observed power-law distributions of neuronal avalanche sizes that is independent of criticality. This opinion paper will describe this phenomenon in more detail, framing it within the existing literature and distinguishing it from related processes which require criticality (e.g., {\bf self-organized criticality (SOC)}).

In the remainder of this paper, the main pieces of evidence provided in support of the criticality hypothesis will be presented. Then, this memory-induced long-range order (MILRO) mechanism will be discussed, which the authors claim offers a more straightforward explanation of the existing experimental data.

\subsection{A clarification on criticality, scale invariance, and long-range order}

Before continuing, it must be emphasized that the term ``criticality'' is not always used consistently. This point is now well understood in the literature~\cite{10.1162/netn_a_00269, how_critical_is_brain_criticality}, but reiterating it once more will provide some useful context. While criticality does have a rigorous definition in physics (as explained above), it is sometimes conflated with the notion of {\it scale invariance}. As the name implies, a system is said to be scale-invariant if it has no characteristic length scale. In other words, zooming in or out on a scale-invariant system leaves its structure unchanged.

Additionally, and almost by definition, systems with no characteristic size can have no ``cut-off,'' meaning power laws are to be expected. However, although power-law distributions are hallmarks of scale invariance, they are not sufficient to prove that a system is critical. Criticality implies scale invariance~\cite{tauber2014critical} but the inverse is not necessarily true. There are various other ways to generate scale invariance in the absence of criticality, as several studies have amply demonstrated~\cite{PhysRevE.66.067103, Mitzenmacher01012004, touboul2017power}.

Lastly, there is the yet-weaker notion of {\bf long-range order}, or {\it long-range correlations}, which is not always related to criticality. Systems with long-range order have elements that indirectly depend on, or {\it correlate} with, other elements quite strongly over long distances, much longer than their interaction length. Physicists might describe this type of system as {\it strongly coupled}. This does not necessarily mean that individual elements {\it explicitly} depend on all other elements in the systems, but rather that their {\it implicit dependence} is significant enough that it cannot be ignored. The presence of long-range correlations is also necessary, but not sufficient, for criticality.

\section{Addressing Some Arguments in Favor of The Criticality Hypothesis} 

\subsection{Properties of power-law exponents}

Since power laws alone are not sufficient to imply criticality, researchers have gone beyond the mere existence of power laws to study the values of and mathematical relationships between different power-law exponents. In particular, one can study neuronal avalanche distributions as functions of size $s$ and duration $T$ independently. Each of these distributions will decay with its own, unique power-law exponent. Additionally, one can relate average avalanche size (denoted by $\langle s \rangle$) to avalanche duration, yielding a third exponent. These relations are given below:

\begin{equation}
P(s) \sim s^{- \alpha_s} ,\quad\qquad P(T) \sim T^{- \alpha_T} ,\quad\qquad \langle s \rangle (T) \sim T^\gamma .
\end{equation}

Sethna and other researchers~\cite{Sethna2001, sethna2004random} have shown that, as long as a system is scale-invariant, the relationship between exponents $\frac{\alpha_T - 1}{\alpha_s - 1} = \gamma$ will hold. These scaling relationships enable data to be collapsed or re-shaped, which can help eliminate finite-size artifacts and further verify true scale-invariance. This is true {\it regardless} of whether or not the underlying mechanism for these power laws is criticality~\cite{fisher1972scaling}.

Nonetheless, this result has been used as a test of criticality in neural systems~\cite{PhysRevLett.108.208102, fontenele2019criticality}, even though, as has already been emphasized, {\it scale invariance is necessary, but not sufficient, for criticality}. Several studies have demonstrated this by explicitly showing that this relationship holds in {\it non}-critical systems as well~\cite{touboul2017power, PRXLife.3.013013, sun2025}. This misinterpretation may exist because this result was originally provided in the context of critical systems, or because these power-law exponents are often given the moniker of ``critical'' exponents.

\subsection{Maximal information processing}

Furthermore, several studies have argued that criticality optimizes properties like information transmission~\cite{langton1990computation, 10.1098/rsta.2007.2092, ramo2007measures, shew2011information}, storage~\cite{langton1990computation}, computational power~\cite{bertschinger2004edge}, and more~\cite{doi:10.1073/pnas.2302730121, doi:10.1073/pnas.2418218122}. However, these studies generally emphasize the coincidence of scale-invariant avalanches (and other hallmarks of long-range order) with heightened performance. Crucially, {\it they do not require criticality}, only power-law correlations. As such, these properties could well be optimized in systems with long-range order that are not necessarily critical.

Additionally, it is unclear whether the ``optimal'' operation of several brain processing functions is even biologically necessary. In fact, it has been proposed~\cite{WILTING2019105} that maximizing these processing functions simultaneously may not be required. Instead, perhaps performance must merely exceed some sufficient threshold for a given task. Beyond this point, it is feasible that the more relevant biological constraint would be minimizing energy consumption.

One concrete piece of evidence that the brain may optimize information transfer comes from Beggs and Plenz's seminal work~\cite{Beggs11167}, where they detected a power-law exponent of $-3/2$ in their avalanche size distributions. They also observed that heightened electrical activity in one region would, on average, lead to activity in exactly one neighboring region. Both of these results are consistent with what is known as a critical {\bf branching process}, a mathematical framework which models how information is transferred in specific types of networks. They then argued that this ``critical'' state optimizes information transfer by simultaneously enabling effective signal propagation while preventing spiking from being so prevalent that the information of where a signal originated from is washed out.

However, since this initial result, several studies, both experimental~\cite{priesemann2009subsampling, hahn2010neuronal, yaghoubi2018neuronal, ma2019cortical} and numerical~\cite{10.1371/journal.pcbi.1000846, orlandi2013noise, di2018landau, dalla2019modeling, sun2025} have observed power-law exponents which differ from $-3/2$. Furthermore, the value of these exponents often depends on the binning procedure used to collect avalanches, including bin size and activation threshold~\cite{pasquale2008self, touboul2010can, priesemann2013neuronal, PRXLife.2.023008, sun2025}. It is thus unclear how relevant the precise value of these exponents is.

\section{Possible Explanations for Experimental Data}

With this discussion in mind, one point appears irrefutable: power laws are ubiquitous in cortical dynamics. So, what could cause them to appear in these systems?

\subsection{Self-organized criticality}

A dominant paradigm in physics~\cite{PhysRevLett.59.381, watkins2016}, geology~\cite{bak1989earthquakes}, economics~\cite{scheinkman1994self}, and countless other fields is the notion of {\it self-organized criticality} (SOC). This is a process whereby a system is attracted {\it dynamically} in phase diagram to a point of continuous (i.e., second-order) phase transition. This behavior requires both slow driving and fast relaxation, effectively ``tuning'' the system near a critical point where its correlations then become long-ranged.

Several studies claim to have observed such tuning in cortical systems and have attributed it to self-organization~\cite{ma2019cortical, pasquale2008self, 10.3389/fnsys.2014.00166, PhysRevE.101.022303, synapse_SOC, sensory_input_adaptation, criticality_meets_learning}. This could explain the frequency at which power laws and other indicators of long-range order are detected in cortical systems. Even if second-order phase boundaries, where power laws naturally arise, occur in an infinitesimal region in the phase diagram, a dynamical tuning process that pushes the system near this boundary would, in principle, be sufficient to produce long-range correlations.

However, SOC implicitly assumes {\it the existence} of a critical point in the space of parameters (which is dynamically attractive). Confirming this would require some {\bf renormalization group} (RG) analysis~\cite{wilson1983renormalization, kardar2007statistical}, a mathematical procedure which iteratively coarse-grains out small-scale degrees of freedom, shifts model parameters as part of a ``RG flow,'' and identifies critical points as fixed points of this RG flow. While this approach has been used phenomenologically in several neuroscientific works~\cite{criticality_setpoint_review, agrawal2019scale, edge_of_instability_scaling_mouse, ponce2023critical, PRXLife.2.023008}, tackling this analytically in the context of cortical systems has generally eluded researchers thus far. In one study which did perform an analytical RG calculation~\cite{sooter2024cortex}, it was unclear to the authors how well their model corresponded to realistic neural systems. Further study would provide more concrete evidence about the nature of critical points in models of cortical systems.

Perhaps most interestingly, a few studies~\cite{PhysRevE.101.022303, sun2025} that have studied a large region in parameter space by sweeping over a model parameter have observed scale-invariance over a wide regime. Since there is not yet an analytical understanding of these systems' space of parameters, the authors argue that it is simpler to assume these points lie within a {\it phase} with long-range correlations rather than in the vicinity of a yet-detected critical point.

\subsection{Memory-induced long-range order}

If there is indeed an entire phase of long-range order in neural activity, what generates it? Where does it come from? Recent work has indicated that the underlying cause may be the presence of {\it memory} in the system.

Here, ``memory'' does not signify information storage in the traditional sense. Rather, it refers to {\it time non-locality}. To have time non-locality in a system, such as a collection of neurons, means that a system's response to some stimulus depends on its entire history rather than its instantaneous state~\cite{Kubo_1957}. All physical systems possess some degree of time non-locality (no system can respond to a stimulus instantaneously). However, it is especially evident in systems with two types of coupled variables (degrees of freedom). Suppose one of these types evolves faster than the other. Their coupled equations might look something like this:

\begin{align}
  \dot{\rho} &= F_{\rho}(\rho, R) ,\label{rho}\\
  \dot{R} &= \frac{1}{\tau_D}F_{R}(\rho, R) ,\label{R}
\end{align}

\noindent where the functions $F_\rho$ and $F_R$ govern the two variables' dynamics, and $\tau_D \gg 1$ is the ratio of their characteristic timescales. One can call the faster variables ($\rho$) the ``primary'' variables and the slower ones ($R$) the ``memory'' variables. This is because, assuming these types of variables depend on one another, the slower ones will act as a source of memory for the primary ones. This is precisely because they change more slowly than the other variables, so their state at a given time provides a record of the primary variables' past dynamics.

Consider this framework in the context of neural activity. The primary variables are the levels of electrical activity in neuronal populations, which are the easiest observables to experimentally access. These neural activities couple to several slower variables, most notably the availability of metabolic and synaptic resources (e.g., ATP, vesicle availability, and ion-gradient support) at each neuron (see Fig.~\ref{brain_fig}). Indeed, many widely used models~\cite{di2018landau} of neural activity incorporate precisely these two types of variables: activities and resources.

These resource variables span multiple biological timescales: synaptic vesicle pool recovery~\cite{markram1996redistribution, tsodyks1997neural, zucker2002short} and Na$^+$/K$^+$-ATPase-driven ion-gradient restoration~\cite{heinemann1975undershoots, rasmussen2019cortex} on seconds, intracellular ATP transients on seconds to minutes~\cite{natsubori2020intracellular}, and homeostatic synaptic scaling on hours to days~\cite{ma2019cortical}, each now accessible via established readouts such as Intensity-based ATP-Sensing Fluorescent Reporter 2 imaging~\cite{marvin2024iatpsnfr2}, Fei-Mao-dye and vGlut-pHluorin (vesicular glutamate transporter fused to pHluorin) labeling~\cite{doi:10.1126/science.1553547, miesenbock1998visualizing, voglmaier2006distinct}, and ion-selective microelectrodes~\cite{sykova2012ion}. MILRO requires only that these slow processes be much slower than the fast spiking dynamics, not that they share a single characteristic time, so the predicted phenomenology is robust across this hierarchy.

It is clear that the amount of resources and level of electrical activity must depend on each other; neurons cannot spike without resources, and spiking will temporarily deplete available resources. Furthermore, since spiking occurs much more rapidly than the rate of growth or decay of the resources, the resources {\it do} act as a source of memory in the system. Simply by extracting the amount of resources available to a neuron at a point in time, one can infer whether or not it has been a long time since that neuron spiked. Thus, the memory variables provide a record of the neurons' recent dynamics.

Crucially, interaction between these two types of variables does yield long-range order in the neuronal variables, and in some cases scale invariance. Sipling {\it et al.}~\cite{sipling2025memory} demonstrated this numerically and provided analytical 
arguments showing that memory in coupled variables {\it can} give rise to long-range order in the fast variables. This analysis {\it does not} require proximity to criticality, distinguishing this phenomenon from SOC, even though both mechanisms require timescale separation. Furthermore, studies of neuronal activity that claim criticality frequently feature a slow dynamical resource variable~\cite{synapse_SOC, PhysRevE.101.022303} or slow sensory input~\cite{sensory_input_adaptation} coupled to the fast spiking neurons. Models of excitatory and inhibitory neurons~\cite{criticality_meets_learning} also feature timescale separation between the two species. This makes  memory-induced long-range order equally compatible with the experimental data, on this front, as SOC.

\subsection{How then can criticality be distinguished from a phase of long-range order?}

In essence, these two mechanisms differ in their explanation of the underlying system's phase diagram. The criticality hypothesis asserts that the brain exhibits long-range correlations within a narrow band of phase transition, which a SOC process ``pushes'' the system towards. MILRO contends that the wide regime of long-range order is part of a distinct phase not associated with criticality. Although the authors find the MILRO hypothesis more physically plausible, largely due to the significant breadth in model parameters over which long-range order appears to be preserved~\cite{PhysRevE.101.022303, sun2025}, the best way to invalidate one or the other would be through further experimentation.

\begin{figure*}
\centering
\includegraphics[width=\linewidth]{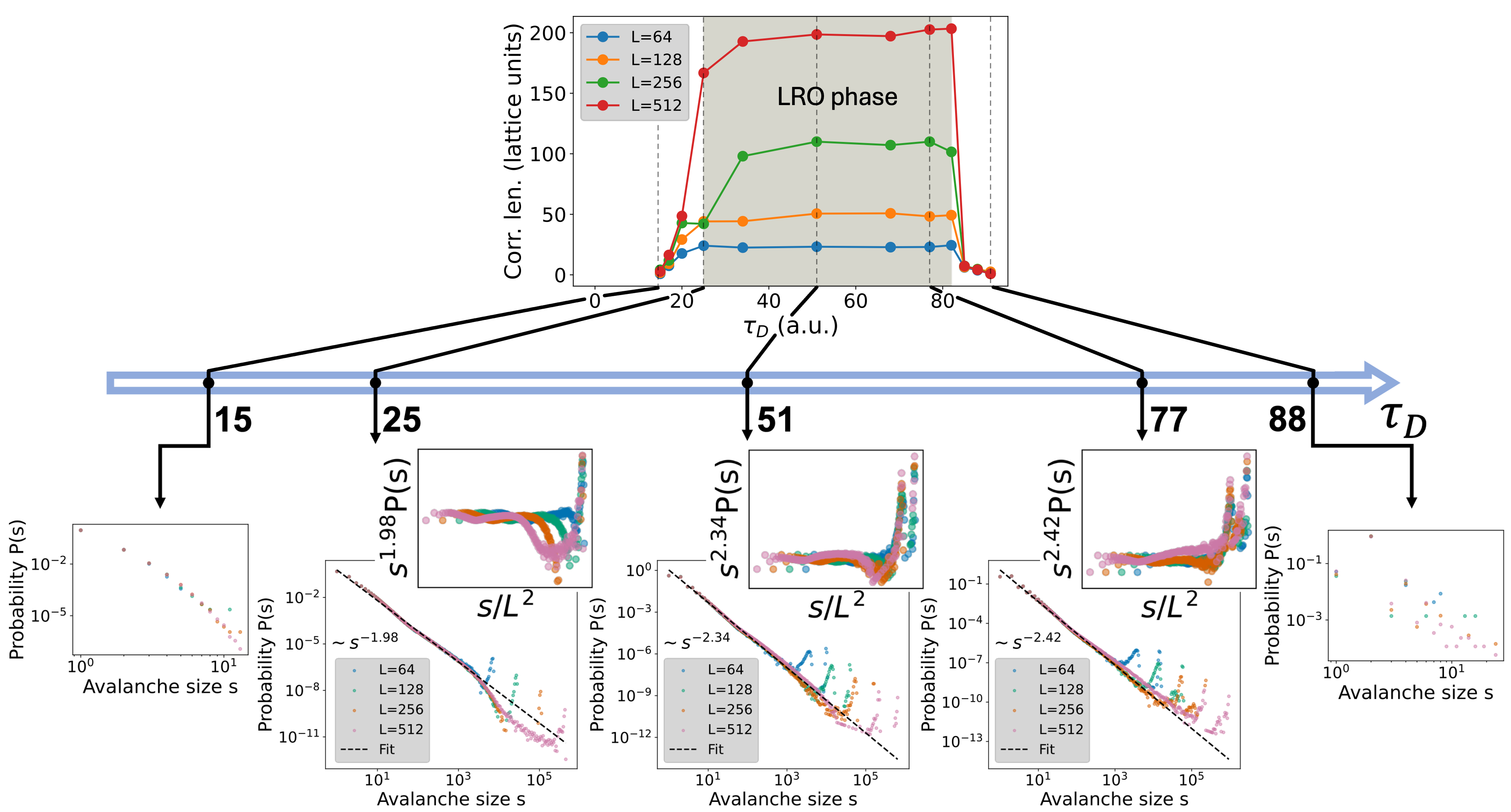}
\caption{A plot of {\bf correlation length} alongside a set of avalanche size distributions, as reported in Sun {\it et al.}~\cite{sun2025}, obtained from modified Wilson-Cowan equations for neural activity~\cite{wilson1972excitatory,di2018landau} solved on a square lattice of side $L$. In this visualization, the parameter $\tau_D$ is varied, which represents the timescale over which the available resources decay. For a wide range of $\tau_D$ ($25 \lesssim \tau_D \lesssim 82$), avalanche distributions well fitted by power laws are observed, and the correlation length approaches the bounds of the lattice, indicating scale-invariant behavior. Outside of this range, regions of minimal ($\tau_D \lesssim 25$) and perpetual ($\tau_D  \gtrsim 82$) activity are observed, where the correlation length shrinks significantly. This suggests that the long-range-ordered state, which several studies have found the brain to operate at, may be part of a wider phase of long-range order. This is difficult to reconcile with the notion that the brain must operate at or near a single critical point of a phase transition. Here, the correlation length is defined as the average distance between two units in the same avalanche. Visualization adapted from Sun {\it et al.}~\cite{sun2025}}
\label{sun_phase_LRO_fig}
\end{figure*}

Specifically, the two pictures make distinct empirical predictions when the slow-resource timescale $\tau_D$ is varied in a controlled way (see Fig.~\ref{sun_phase_LRO_fig}). Most notably, when parameters are altered to move a system away from criticality, avalanche distributions quickly become exponentially decaying~\cite{PhysRevResearch.3.013107, stauffer2018introduction}. Instead, within a MILRO phase, power laws persist in avalanche distributions across the entire phase~\cite{sipling2025memory}.

Some experimental observations cited as evidence for SOC are also compatible with MILRO. For example, Ma {\it et al.}~\cite{ma2019cortical} report that monocular deprivation transiently disrupts power-law avalanche statistics in the visual cortex, which then recover over the course of approximately a day. The MILRO phase is itself ``attractive'' in a different sense: long-range correlations only emerge after a finite time during which memory accumulates in the slow variables~\cite{sipling2025memory}. A dramatic perturbation of the system disrupts these slow variables, so a finite recovery interval before power-law statistics return is expected even in the absence of dynamical attraction toward a critical point.

Specifically, since MILRO relies upon {\it two} types of coupled variables, it would be ideal to study the degree of resource availability in the brain during normal spiking dynamics. Recent advances in genetically encoded sensors~\cite{marvin2024iatpsnfr2}, optical imaging techniques~\cite{li2011concurrent,bower2021high}, and simultaneous metabolic-activity measurements~\cite{bose2024simultaneous} make such experiments increasingly feasible.

Past studies provide a clue as to what experimental techniques could facilitate this. In their 2005 paper, Haldeman and Beggs~\cite{haldeman2005critical} demonstrated that avalanche statistics {\it did} change when their cortical samples were chemically altered to encourage excitation (more spiking) or inhibition (less spiking). As expected, this yielded avalanche size distributions that deviated somewhat from power-law decays, often called ``super-critical'' and ``sub-critical,'' respectively. This has been used as evidence that the central point is critical~\cite{10.3389/fphys.2012.00163}.

However, a similar approach could be used {\it in vitro} to more thoroughly explore the phase diagram along the $\tau_D$ (memory timescale) direction. Traditional microelectrode arrays could be paired with optical devices to monitor vesicle recycling rates~\cite{doi:10.1126/science.1553547, synaptic_slice_vis}. Since vesicles play a key role in enabling neurotransmitters to propagate, thus gating the flow of electrical activity in neural systems, and are recycled over a much longer timescale than the spiking time of individual neurons, their recycling rate could act as an effective resource frequency. Once a baseline activity and resource timescale are established, 2-Deoxyglucose (2-DG), which inhibits glycolysis and thereby reduces glycolytic ATP production, could be then introduced. By slowing down ATP production, processes that rely upon ATP availability (like vesicle recovery) would slow down as well. Effectively, titrating 2-DG into neuronal cultures {\it in vitro} could slowly alter the effective resource timescale, thereby exploring a substantial line in the phase diagram. It is also possible that, in the near future, even more sophisticated tools could be applied similarly to measure neural activities~\cite{doi:10.1126/science.abf4588} and intracellular ATP~\cite{marvin2024iatpsnfr2} in higher resolution.

Furthermore, new research indicates that long-range order can even be ``dragged'' across layers in multi-layered media~\cite{zhang2026memory}. Volumetric neural imaging techniques now exist~\cite{rabut20194d, demas2021high, gutierrez2022unique} that could probe this behavior in the neocortex, which itself is organized into distinct layers~\cite{felleman1991distributed, douglas2004neuronal, harris2015neocortical, mejias2016feedforward}.

Hopefully, this opinion paper will spark interest in pursuing these experiments.

\section{CONCLUDING REMARKS}

In his 2022 paper~\cite{10.3389/fncom.2022.703865}, Beggs addresses recent skepticism of the criticality hypothesis, particularly that of Destexhe and Touboul~\cite{touboul2017power}. He claims that their counterexamples, which satisfy many of the criteria of critical systems without being critical themselves, are insufficient because they do not possess three key features that must exist in experimental neural systems. His point is that, in real neural systems, any signatures of criticality must:

1. Emerge due to collective behavior,

2. Possess long-range temporal correlations, and

3. Support information processing.

Memory-induced long-range order {\it does} satisfy each of these physically motivated requirements. First, the long-range order in these systems emerges {\it precisely} due to collective behavior in neural activity, which is induced by the presence of memory as provided by the resources. Additionally, since memory is necessarily present in this system, which inherently requires variables to correlate strongly over long durations, the existence of long-range correlations {\it in time} is also expected. This has been shown explicitly in the work of Sun {\it et al.}~\cite{sun2025} by visualizing avalanche {\it duration} distributions, which also follow power-law decays. The first two points then follow directly from the definition of memory-induced long-range order. Of course, SOC could give rise to this behavior as well, but it requires further assumptions about the underlying system's phase diagram than MILRO.

In regard to Beggs' final point, recent computational work has indicated that long-range order alone may be sufficient for tasks requiring high computational performance. A new computational paradigm known as memcomputing~\cite{di2018perspective, di2022memcomputing} has demonstrated that long-range order induced by memory, when applied to solving hard computational problems, can outperform state-of-the-art algorithms at various benchmarks, supporting the claim that memory-driven collective modes can be robust without fine tuning. Furthermore, recent studies have argued that the brain may exist in different dynamical states based on the external state or relevant computational task~\cite{doi:10.1126/sciadv.adj9303, w49n-2vz8, PRXLife.2.023008}, making the nature of the alleged critical state more ambiguous in these situations.

Nevertheless, much of the recent work intending to demonstrate that the brain operates at criticality does not always distinguish between criticality and long-range order or scale invariance. Studies displaying long-range correlated avalanche distributions, enhanced computational performance, or even a wide dynamic range~\cite{optimal_dynamical_range, Shew15595} can all be explained by {\it either} behavior. Without concrete evidence for the existence of a critical state, the authors argue that the brain may naturally operate in an entire phase of (memory-induced) long-range order. Other studies have argued how scale-invariance could arise in cortical dynamics away from criticality (via neutral drifts~\cite{PhysRevResearch.3.013107}, heavy-tailed connectivity~\cite{Shi2025.08.30.673281}, stochasticity~\cite{PRXLife.3.013013}, etc.), but to the authors' knowledge, an analytical connection to time non-locality has not yet been made. The experimental fact that resources in the brain, which promote neural activity, act as a source of memory, naturally leads to such an explanation without requiring further evidence indicating criticality.

\section{GLOSSARY}

\noindent{\bf Neuron} - A nerve cell that transmits information through electrical and chemical signals.

\noindent{\bf Spike} - An electrical event in a neuron characterized by a rapid increase in voltage across a neuron's cell membrane. These electrical impulses can be transmitted between neurons, enabling communication. This event is also known as an {\it action potential}.

\noindent{\bf Synapse} - The region between two neurons where chemicals produced by spikes in electrical activity can be transferred.

\noindent{\bf Neurotransmitter} - A chemical released due to a spiking event that enables communication between neurons. Neurotransmitters travel across the synapse and can either promote (excite) or hinder (inhibit) the passage of electrical impulses.

\noindent{\bf Neuronal Avalanche} - A collection of electrical events in neurons that occur in rapid succession. Avalanches can involve anywhere between one and all neurons in a system.

\noindent{\bf Criticality Hypothesis} - The hypothesis that the brain operates near criticality, and that at this point, several information processing functions operate optimally due to the collective interactions between neurons~\cite{10.1098/rsta.2007.2092, PhysRevLett.108.208102, 10.1162/netn_a_00269}.

\noindent{\bf Criticality} - A boundary of a continuous (second-order) phase transition separating two or more distinct phases or states. For example, a ferromagnet at the Curie temperature, where spontaneous magnetization disappears and the system becomes paramagnetic, is at criticality.

\noindent{\bf Phase Transition} - The physical transformation of a system between distinct states with unique properties, often through the variation of a single parameter (i.e., a control parameter). For example, a ferromagnet can transition to a paramagnetic state if its temperature is increased.

\noindent{\bf Memory} - A general property in physical systems in which the response of a system to some stimulus depends on the full history of its past dynamics. This is also known as {\it time non-locality}. In this work, the word ``memory'' is used exclusively in this context, rather than in the context of the storage or retrieval of information.

\noindent{\bf Self-Organized Criticality (SOC)} - A process whereby systems are attracted {\it dynamically} in phase diagram to a point of continuous phase transition. This phenomenon requires a separation of timescales, including both slow external driving and fast relaxation~\cite{PhysRevLett.59.381, watkins2016}.

\noindent{\bf Long-Range Order} - The physical property characterized by a slow decay in correlation between units of a system as distance is increased. In other words, units in systems with long-range order more significantly correlate their motion with their distant neighbors than in systems without it. Equivalently, such a system could be said to have {\it long-range correlations}.

\noindent{\bf Branching Process} - A mathematical model which describes how information can be transferred in networks. These processes can be further categorized by their branching parameter, which counts how many neighboring nodes an individual node transfers information to, on average. ``Critical'' branching processes have a branching parameter of one, {\it although they are not necessarily associated with criticality from a phase diagram perspective}.

\noindent{\bf Renormalization Group} - A coarse-graining procedure which integrates out short-distance/high-energy degrees of freedom as part of the ``RG flow.'' This technique provides a more rigorous understanding of the phase diagram structure of a particular model.

\noindent{\bf Correlation Length} - The characteristic length scale over which fluctuations in variables are correlated. Scale-invariant systems have a correlation length that diverges as the system size is increased.

\section{ACKNOWLEDGMENTS}

This work was funded by the National Science Foundation via grant No. ECCS-2229880. M.D. also acknowledges funding by the Alexander von Humboldt Stiftung through the Humboldt Research Award. We thank Val\'erian Demaurex for useful comments on our manuscript.

\section{AUTHOR CONTRIBUTIONS}

M.D. suggested and supervised the work. C. S. wrote the original draft of the manuscript. All authors contributed to the review and editing of the final version. All authors have read and approved the final manuscript.

\section{DECLARATION OF INTERESTS}

The authors declare no competing interests.

\bibliography{references}

\bigskip

\end{document}